\documentclass[conference]{IEEEtran}

\usepackage{algorithm}

\usepackage[shortlabels]{enumitem}
\usepackage{balance}
\usepackage{booktabs}
\usepackage{todonotes}
\usepackage{epsfig}
\usepackage{graphicx}
\usepackage{amsmath}
\usepackage{bm}
\usepackage{lipsum}
\usepackage{subfig}
\usepackage{color}

\usepackage[normalem]{ulem}
\usepackage{bigstrut}
\usepackage{listings}
\usepackage{textcomp}
\usepackage{eso-pic}
\usepackage{tikz}
\usepackage{xspace}
\usepackage{fancyhdr}
\usepackage{mathtools}
\usepackage{soul}
\usepackage{tipa}
\usepackage{textgreek}
\usepackage{siunitx}
\usepackage[nocompress]{cite}
\usepackage{amssymb}
\usepackage{pifont}
\usepackage{xurl}

\usepackage[bookmarks=true,breaklinks=true,letterpaper=true,colorlinks,linkcolor=blue,citecolor=magenta,urlcolor=blue]{hyperref}

\def\BibTeX{{\rm B\kern-.05em{\sc i\kern-.025em b}\kern-.08em
    T\kern-.1667em\lower.7ex\hbox{E}\kern-.125emX}}

\title{BlissCam: \underline{B}oosting Eye Tracking Efficiency with \underline{L}earned \underline{I}n-Sensor \underline{S}parse \underline{S}ampling}

\makeatletter
\newcommand{\linebreakand}{%
  \end{@IEEEauthorhalign}
  \hfill\mbox{}\par
  \mbox{}\hfill\begin{@IEEEauthorhalign}
}
\makeatother

\author{\IEEEauthorblockN{Yu Feng\textsuperscript{*}\textsuperscript{$\clubsuit$}}
\IEEEauthorblockA{Shanghai Jiao Tong University\\University of Rochester\\y-feng@sjtu.edu.cn}
\and
\IEEEauthorblockN{Tianrui Ma\textsuperscript{*}}
\IEEEauthorblockA{Washington University in St. Louis\\tianrui.ma@wustl.edu}
\linebreakand
\IEEEauthorblockN{Yuhao Zhu\textsuperscript{\#}}
\IEEEauthorblockA{University of Rochester\\yzhu@rochester.edu}
\and
\IEEEauthorblockN{Xuan Zhang\textsuperscript{\#}}
\IEEEauthorblockA{Northeastern University\\xuan.zhang@northeastern.edu}
}

\newcommand*\circled[2]{\tikz[baseline=(char.base)]{
            \node[shape=circle,fill=black,inner sep=1pt] (char) {\textcolor{#1}{{\footnotesize #2}}};}}

\ifx\figurename\undefined \def\figurename{Figure}\fi
\renewcommand{\figurename}{Fig.}
\renewcommand{\paragraph}[1]{\textbf{#1} }

\newcommand{\Sect}[1]{Sec.~\ref{#1}}
\newcommand{\Fig}[1]{Fig.~\ref{#1}}
\newcommand{\Tbl}[1]{Tbl.~\ref{#1}}
\newcommand{\Eqn}[1]{Eqn.~\ref{#1}}

\newcommand{\specialcell}[2][c]{\begin{tabular}[#1]{@{}c@{}}#2\end{tabular}}

\newcommand{\proj}{\textsc{BlissCam}\xspace}
\newcommand{\mode}[1]{\underline{\textsc{#1}}\xspace}

\newcommand{\no}[1]{#1}
\renewcommand{\hl}[1]{#1}
\newcommand{\RNum}[1]{\uppercase\expandafter{\romannumeral #1\relax}}

\newcommand{\tianrui}[1]{{\color{magenta} [Tianrui: #1]}}

\graphicspath{{figs/}}

\begin{document}
\maketitle

\begingroup\renewcommand\thefootnote{$\clubsuit$}
\footnotetext{Work done while at University of Rochester}
\endgroup
\begingroup\renewcommand\thefootnote{*}
\footnotetext{Equal contribution}
\endgroup
\begingroup\renewcommand\thefootnote{\#}
\footnotetext{Corresponding authors}
\endgroup

\thispagestyle{plain}
\pagestyle{plain}

\setcounter{page}{1}

\begin{abstract}

Eye tracking is becoming an increasingly important task domain in emerging computing platforms such as Augmented/Virtual Reality (AR/VR).
Today's eye tracking system suffers from long end-to-end tracking latency and can easily eat up half of the power budget of a mobile VR device.
Most existing optimization efforts exclusively focus on the computation pipeline by optimizing the algorithm and/or designing dedicated accelerators while largely ignoring the front-end of any eye tracking pipeline: the image sensor.
This paper makes a case for co-designing the imaging system with the computing system.

In particular, we propose the notion of ``in-sensor sparse sampling'', whereby the pixels are drastically downsampled (by 20$\times$) within the sensor.
Such in-sensor sampling enhances the overall tracking efficiency by significantly reducing 1) the power consumption of the sensor readout chain and sensor-host communication interfaces, two major power contributors, and 2) the work done on the host, which receives and operates on far fewer pixels.
With careful reuse of existing pixel circuitry, our proposed \proj requires little hardware augmentation to support the in-sensor operations.
Our synthesis results show up to 8.2 $\times$ energy reduction and 1.4 $\times$ latency reduction over existing eye tracking pipelines.

\end{abstract}


\section{Introduction}
\label{sec:intro}

Eye tracking provides a fundamental utility in many fields, ranging from medical studies~\cite{brunye2019review, borys2017eye} and human-machine interaction~\cite{mania2021gaze, strandvall2009eye, jacob2003eye, chandra2015eye, mathur2021dynamic} to augmented/virtual reality (AR/VR) and spatial computing~\cite{itoh2014interaction, plopski2016automated, hu2020gaze, clay2019eye, whitmire2016eyecontact, patney2016perceptually, zhang2017eye}.
Accurately and efficiently tracking eye gazes play an important role in understanding human cognition~\cite{wang2021multi}, enabling gaze-based human-machine interactions~\cite{sidenmark2022weighted}, and improving communication and computation efficiency of AR/VR systems~\cite{Patney:2016:TFR, duinkharjav2022color, Chakravarthula:2021:GCR, hou2023architecting, feng2019asv}.

Despite its essentiality, eye tracking is known to be slow and power-hungry~\cite{mayberry2014ishadow}.
In our measurement of commercial eye trackers (e.g., HTC Vive Pro Eyes and Tobii), its latency is usually in excess of 15 \si{\milli\second}, enough to introduce visual artifacts (e.g., in gaze-contingent rendering), and its always-on status constantly consumes over 2 \si{\watt}~\cite{tobii_eyetracker}, eating up half of the power budget of a typical VR system~\cite{rift_power, quest2_power, quest3_power}.

Most of today's efforts in optimizing eye tracking focus on the algorithm pipeline, either by 
optimizing the tracking algorithms~\cite{feng2022real, kothari2021ellseg, chaudhary2019ritnet, angelopoulos2020event, kim2016real} or by designing dedicated hardware accelerators~\cite{you2022eyecod, zhao2022flatcam}, while largely ignoring the indispensable front-end of any eye tracking pipeline: the image sensor, which generates near-eye images for the tracking algorithms to consume.

\paragraph{Rationale.}
This paper makes a case for jointly designing the imaging system and the tracking algorithm to significantly reduce energy consumption while satisfying the stringent tracking latency requirement.
In particular, we propose the notion of ``in-sensor sparse sampling'', whereby the pixels are drastically down-sampled (retaining only about 5\% of the pixels) \textit{within the sensor}.
The downstream eye tracking algorithm is carefully co-designed to be robust and take advantage of the sparse inputs.

Such sensor/algorithm co-design offers two unique opportunities.
First, we can optimize a previously untapped system component with significant power and latency implications, namely the image sensor.
Modern image sensors, along with their communication interfaces, are power hungry; they consume power upwards of a few Watts, making up half of the eye tracking power.
The sensor power is dominated by the analog readout chain and the sensor-host data transfer, both of which are decreased with in-sensor data reduction.
Second, with sparsely sampled sensor data, the host eye tracking algorithm receives and, thus, operates on far few pixels, further reducing the tracking latency and energy consumption.

\paragraph{Contributions.}
We make both algorithmic and architectural contributions.
Algorithm-wise, we show how to design the image sampling algorithm to reduce tracking latency and energy without hurting the accuracy.
We observe that the background in near-eye images is stationary and the only moving pixels in an image contribute to the gaze result.
This observation leads to a two-stage sampling algorithm, which first detects the moving parts of an image as the Region-of-Interest (ROI) followed by random sampling within the ROI.

We propose an eye tracking algorithm to take advantage of the pixel reduction.
The algorithm is based on Vision Transformers whose accuracy is robust against pixel sparsity and whose cost of computation naturally reduces as the pixel volume reduces.
Critically, both the data sampling algorithm and the eye tracking algorithm are (approximately) differentiable, which allows us to jointly train the in-sensor and off-sensor operations to maximize end-to-end tracking accuracy.

The architectural contribution of the paper is to minimally augment the sensor architecture to support various in-sensor operations.
We base our design on the increasingly popular die-stacked Digital Pixel Sensor (DPS) architecture, where the top layer is the usual pixel array and the bottom layer integrates per-pixel ADC/SRAM and a DNN accelerator~\cite{bong2017low, eki20219, seo20222, hsu20230}.
We show how to map the in-sensor sampling algorithm to the bottom layer by time-multiplexing existing circuit components between the sparse sampling mode and the usual imaging mode.

\paragraph{Results.}
According to results obtained through digital logic synthesis and analog circuit-level simulation, we show that our eye tracking system reduces pixel volume by about 95\%, leading to an 8.2$\times$ energy reduction and a 1.4$\times$ tracking latency reduction compared to existing eye tracking systems, all with little degradation on the tracking accuracy.
Our end-to-end trained in-sensor sampling strategy and eye tracking algorithm consistently outperform baselines and other variants in accuracy across a range of sampling rates, showing the benefits of joint design of in-sensor and off-sensor algorithms.

\hl{In summary, we present a new form of algorithm-hardware co-design, where the hardware spans both the conventional accelerators and, critically, the image sensor.
We expand the research scope from optimizing only for (DNN) accelerators to the end-to-end eye tracking pipeline, which necessarily includes the image sensor.
The key in our work is to optimize the sensor architecture \textit{jointly} with the off-sensor computation — through sparse in-sensor sampling. We hope that the paper can inspire follow-up work on joint sensing-computing system optimization.
Our specific contributions are:}

\begin{itemize}
    \item We analyze the technology trend and pinpoint the system-level bottleneck of today's eye tracking pipeline.
    \item We propose an in-sensor sparse sampling algorithm jointly designed with an off-sensor eye tracking algorithm. They collectively reduce the pixel volume while preserving high gaze prediction accuracy.
    \item \hl{We propose hardware augmentations, both analog (\mbox{\Fig{fig:cis_timing}}) and digital (\mbox{\Fig{fig:roi_select}}), to support in-sensor sparse sampling for the first time. The hardware extensions are \textit{intentionally} kept minimum, enabled by intelligently reusing existing hardware components.}
    \item \hl{We demonstrate a systematic integration of eventification, ROI prediction, sampling, and readout, which serves as a reference design for future stacked image sensors that are increasingly integrating computation capabilities.}
    \item \hl{We propose a new timing design that schedules the hardware components, within and off the sensor, to ensure that the FPS is unaffected by the addition of new computations and hardware components (\mbox{\Fig{fig:pipeline}}).}
    \item Together, we achieve an 8.2$\times$ energy saving and 1.4$\times$ latency reduction compared to existing eye tracking systems with negligible accuracy compromise.
\end{itemize}
\section{Background and Motivation}
\label{sec:motiv}

\begin{figure}[t]
\centering
\includegraphics[width=\columnwidth]{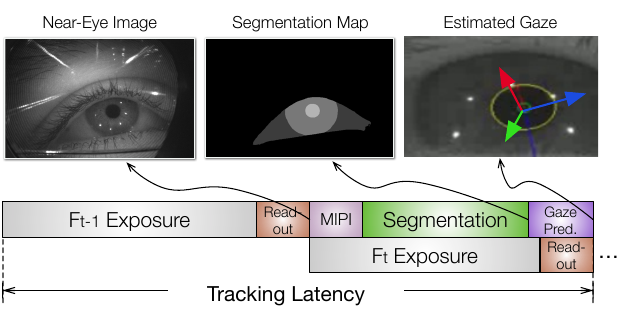}
\caption{A typical eye tracking pipeline, which starts from image sensing (exposure and readout) to obtain an near-eye image, which is transferred to the host processor through the MIPI CSI-2 interface.
The host processor first segments important eye parts, from which the gaze is estimated.
Different frames are overlapped to improve tracking frequency.
Figure not drawn to scale; readout delay is usually three orders of magnitude shorter than the exposure time.}
\label{fig:overview}
\end{figure}

We first review today's mainstream eye-tracking pipeline (\Sect{sec:motiv:eye}) and the basics of image sensors (\Sect{sec:motiv:cis}).
We then discuss the scaling trends of eye tracking technologies in AR/VR, motivating the paper (\Sect{sec:motiv:mot}). 

\subsection{Eye Tracking Basics}
\label{sec:motiv:eye}

The goal of eye tracking is to estimate the user's real-time gaze---a 3D vector indicating where the eye is looking.
It provides a core utility for a variety of human-machine interfaces.
In particular, eye tracking is essential to \hl{next-generation AR/VR systems}, where the rendering is contingent on gaze information~\cite{roth2017quality, kim2018eye} and the user interface (UI) is controlled by gaze~\cite{mania2021gaze, strandvall2009eye, jacob2003eye, chandra2015eye, mathur2021dynamic}.
Apart from AR/VR, eye tracking is also widely used in vision science~\cite{borys2017eye, intoy2020finely, rucci2007miniature}, cognitive study~\cite{rahal2019understanding, semmelmann2018online}, and education~\cite{jarodzka2021eye}.


\paragraph{Eye Tracking Pipeline.} A typical eye tracking pipeline is illustrated in \Fig{fig:overview}.
An image captured by a near-eye camera goes through two stages: eye segmentation, which dissects the foreground eye parts (e.g., pupil, iris, cornea), and gaze prediction, which predicts the gaze from the segmentation map
~\cite{yiu2019deepvog, hennessey2006single, chen20083d, zhu2005eye, li2018etracker, zhang2019evaluation}.
In today's state-of-the-art eye tracking system, the eye segmentation stage is usually performed through Deep Neural Networks (DNNs), whereas the gaze prediction stage employs regression models based on the geometric model of human eyes, making
eye segmentation considerably more time-consuming than gaze estimation.


\paragraph{System Specifications.}
It is shown that the tracking frequency needs to be around 120 Hz with a tracking latency of sub-10 \si{\milli\second} and an accuracy of 0.5-1.0$^\circ$~\cite{smarteyetracker, tobiieyetracker, eyelink, htc_eye_tracker}.
\hl{120 Hz is necessary because humans frequently make rapid eye movements (i.e., saccades) whose speed can be up to 700°/s\mbox{\cite{saccade2009}}, the tracking rate must be high to track such rapid movements.}
Meanwhile, we must work with a tight power envelope available on mobile AR/VR devices (around 3-6 \si{\watt}~\cite{rift_power, quest2_power, quest3_power}) to avoid user discomfort induced by the thermal effect.
The image sensor, host processor, and sensor-host communication all contribute to the power consumption of eye tracking, which we will describe next.



\subsection{Image Sensor Basics}
\label{sec:motiv:cis}

When exposed to light, an image sensor transforms optical signals in the scene to analog signals (using the photoelectric effect~\cite{einstein1905molekularkinetischen}).
The analog signals are converted to digital pixel values through the readout chain (including the ADCs).
\hl{Through the Mobile Industry Processor Interface Camera Serial Interface 2 (MIPI CSI-2)\mbox{\cite{mipi_csi}}}, the pixel values are then transferred to a host processor to undergo further algorithmic processing (e.g., eye tracking).
These operations are necessarily serialized within a frame but can be overlapped across frames.
\Fig{fig:overview} illustrates a typical overlapping between frames, where the next frame can start its exposure while the previous frame's pixels are being transferred out.

\paragraph{Frame Rate.}
A key performance metric of image sensors is the frame rate, quantified by Frames Per Second (FPS).
Ideally, as is the case in \Fig{fig:overview}, the MIPI transfer and host processing delay (of the current frame) is completely hidden by the exposure and the readout delay (of the previous frame).
In this case, the frame rate is limited only by the sum of exposure time and readout delay.
Note that the readout delay (tens of \si{\micro\second}) is usually 3-4 orders of magnitude shorter than the exposure time (tens of \si{\milli\second}). 

\paragraph{Stacked Image Sensors.}
Image sensors today are increasingly integrating advanced \textit{computation} capabilities through 3D die stacking technologies, presenting opportunities for architectural exploration. 
Nowadays almost all mobile image sensors are stacked~\cite{scotttrends}: the pixel array layer that converts photons to analog signals and the processing layer which contains the readout circuitry and the preprocessing image signal processor (ISP) are located on two separate dies.

With the additional stacking dimension, computational image sensors now routinely integrate digital logic (e.g., DNN accelerators~\cite{bong201714, bong2017low, eki20219}) and memories (e.g. DRAMs~\cite{tsugawa2017pixel}, and SRAMs~\cite{seo20222, hsu20230}).
Moreover, stacking allows for heterogeneous integration, where the pixel array layer and the processing layer can each use their respective process node.
A recent survey~\cite{chossatimage} shows that it is common for the processing layer to adopt a process node (e.g., 22 \si{\nano\meter}) that is several generations ahead of the one used by the pixel array layer (e.g., 65 \si{\nano\meter}) in order to accommodate high-density energy-efficient digital processing.

\paragraph{Digital Pixel Sensor.} 
Our paper adopts the stacked Digital Pixel Sensor (DPS), a particular form of image sensor architecture that is gaining popularity~\cite{wuu2022review}.
In DPS, the processing layer has a per-pixel ADC, which inherently supports a global shutter (critical for high-speed capturing) and a per-pixel SRAM to store the digitized pixel value and naturally act as the input buffer to a digital accelerator.
Examples of DPS include the imager prototypes by Meta~\cite{ikeno20224}, Samsung~\cite{seo20222}, OmniVision~\cite{omni_cis}, and Sony~\cite{imx500}.

The recent trend in computational image sensors suggests that it is possible to integrate domain-specific accelerators into the processing layer of a stacked DPS sensor with minimal area overhead.
For instance, under a 22 \si{\nano\meter} process node, our experiment shows that integrating a DNN processor merely introduces 5.8\% of area overhead (\Sect{sec:eval:area}).
Many such designs have been proposed by previous works using Arithmetic Logic Units (ALUs) under a group of pixels~\cite{millet20195500} and CNN processor under the entire pixel array~\cite{eki20219,sharda2023temporal}.

\subsection{Scaling Trends of Eye Tracking Technologies}
\label{sec:motiv:mot}

\begin{figure}[t]
\centering
\includegraphics[width=\columnwidth]{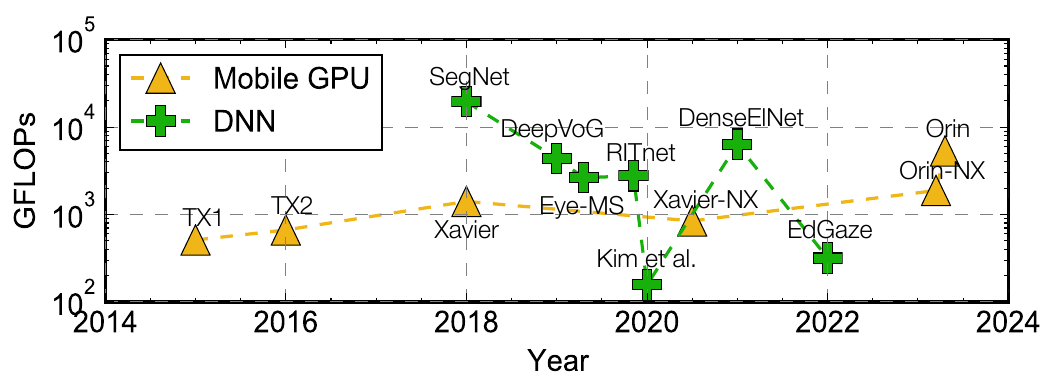}
\caption{The computational capabilities, quantified in GFLOPS, of today's mobile GPUs (using Nvidia Jetson series as examples) vs. the computational demands of state-of-the-art eye tracking algorithms (assuming a tracking rate of 120 FPS).}
\label{fig:comp_trend}
\end{figure}

\paragraph{Tracking Frequency.}
Eye tracking performance is mainly quantified by the tracking frequency: how many gaze estimations can be made in a second.
This paper assumes a 120 \si{\hertz} tracking rate, which is shown to be sufficient for many eye tracking use cases (e.g., AR/VR)~\cite{hooge2022robust, chang2021high} and is on par with commodity eye trackers~\cite{smarteyetracker, tobiieyetracker, eyelink}.

Meeting the tracking rate does not pose any issue as technologies scale.
This is because the speed of the eye tracking algorithm on recent mobile processors with embedded accelerators (e.g., GPUs), is already higher than that of the image sensor's capturing rate.
Thus, the algorithm delay can be hidden by the exposure time (\Fig{fig:overview}).

To quantify this argument, \Fig{fig:comp_trend} compares the Giga Floating Point Operations per second (GFLOPs) of the mobile GPUs on Nvidia Jetson series~\cite{jetson_serial} with the GFLOPs requirement of a set of state-of-the-art eye tracking algorithms operating at 120 Hz~\cite{chaudhary2019ritnet, yiu2019deepvog, kothari2021ellseg, feng2022real}.
The GPUs and algorithms are placed on the $x$-axis based on their release dates.
As mobile computing capabilities (unsurprisingly) increase over the years, eye tracking algorithms also become more efficient.
Hence, the tracking rate requirement can be adequately met by today's mobile GPUs.



\begin{figure}[t]
\centering
\begin{minipage}[t]{0.48\columnwidth}
  \centering
  \includegraphics[width=\columnwidth]{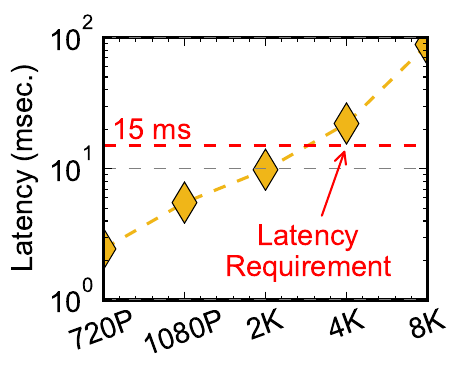}
  \caption{MIPI communicating latency under different image resolutions. The red line shows the eye tracking latency requirement (15 \si{\milli\second}).}
  \label{fig:mipi_trend}
\end{minipage}
\hspace{2pt}
\begin{minipage}[t]{0.48\columnwidth}
  \centering
  \includegraphics[width=\columnwidth]{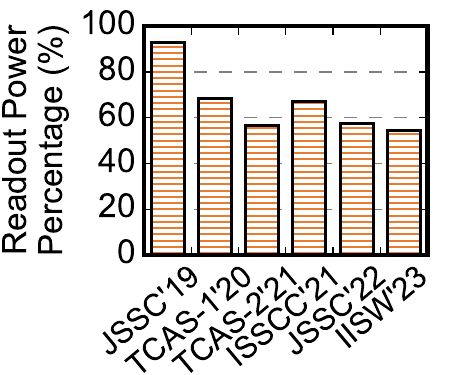}
  \caption{\hl{Percentage of image sensor power attributed to by the readout circuitry; data from six recent sensors}~\cite{park2019640,park2020ultra,park2021cmos,singh202134,seo20222,ikeno2023evolution}.}
  \label{fig:cis_energy_breakdown}
\end{minipage}
\end{figure}

\begin{figure*}[t]
\centering
\includegraphics[width=\textwidth]{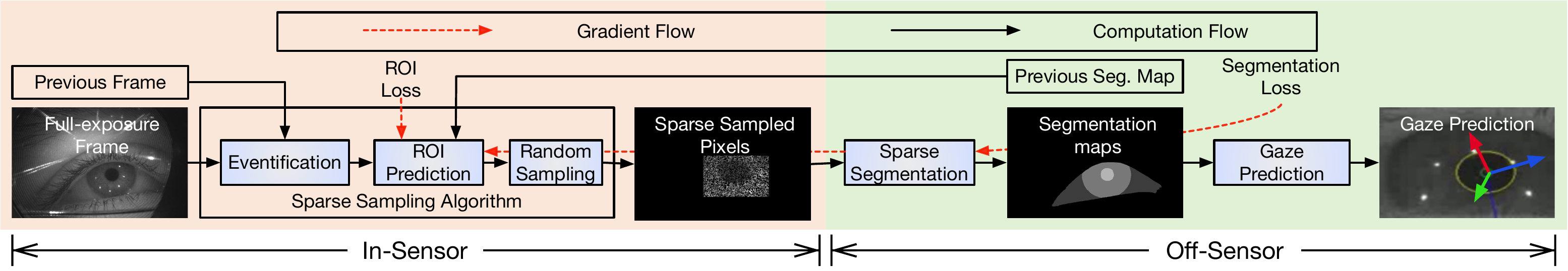}
\caption{Our sparse sampling-based eye tracking pipeline.
Each frame first gets sampled by our sparse sampling algorithm inside the sensor to dramatically reduce the sensor-host data volume (\Sect{sec:algo:ss}); the sampled pixels then go through a sparse eye segmentation algorithm on the host, which is designed to be robust against sparse inputs (\Sect{sec:algo:dnn}).
The ROI prediction algorithm and the sparse segmentation algorithm are jointly trained to maximize end-to-end tracking accuracy (\Sect{sec:algo:train}).
}
\vspace{-5pt}
\label{fig:sparse_sample}
\end{figure*}


\paragraph{Tracking Latency.}
While tracking rate is unlikely an issue, tracking latency is.
Tracking latency is the delay between the start of a frame exposure and when the eye tracking algorithm finishes on that frame (\Fig{fig:overview}).
Our measurements of commercial eye trackers (e.g., HTC Vive Pro Eyes~\cite{htc_eye_tracker}) show that the tracking latency is usually in excess of 15 \si{\milli\second}, enough for many gaze-contingent AR/VR systems to report tracking delay as a main cause of user-observable artifacts~\cite{imaoka2020assessing, adhanom2023eye}.
\hl{15 \mbox{\si{\milli\second}} is roughly the end-to-end latency under a 120 Hz tracking rate when capturing and processing are fully overlapped (as shown in \mbox{\Fig{fig:overview}}): 15 \mbox{\si{\milli\second} $\approx \frac{1\si{\second}}{120}\times 2$}.}

Exposure time accounts for a large proportion of the tracking latency. Simply reducing the exposure time, however, has noise implications, because the Signal to Noise Ratio (SNR) of image sensing drops \textit{quadratically} with exposure time~\cite{rowlands2017physics}.
Therefore, the downstream operations must be robust against exposure time changes.
Among other components, MIPI CSI-2 transfer is poised to become a latency bottleneck as image resolution increases in future.
\Fig{fig:mipi_trend} shows the MIPI latency under various image resolutions~\cite{mipi_datarate1, mipi_datarate2}.
As the image resolution increases to 4K, the transmission latency (22 \si{\milli\second}) alone already surpasses the end-to-end latency requirement,
\hl{which means the MIPI latency can not be hidden by the processing of the next frame.}


\no{This paper will demonstrate techniques that reduce the MIPI transfer latency and, consequently, the subsequent eye tracking algorithm latency while being robust against (even drastic) changes to the exposure time.}

\paragraph{Power Consumption.}
Eye tracking power consumption is known to be high.
Two recent eye tracking algorithms, RITnet and EdGaze, consume 2.3 \si{\watt} and 1.9 \si{\watt}, respectively, on a mobile Volta GPU~\cite{xaviersoc}.
Apart from the computation power, the power consumption of image sensors has been steadily increasing.
Recent high-speed (120~FPS) image sensors routinely consume hundreds of milliwatts~\cite{kwon2020low,canon5mp} or even a few \si{\watt}s~\cite{mstsensor,ovb0a},
taking 10-60\% of the total power budget of a typical standalone VR device (around 3-6\si{\watt}~\cite{rift_power, quest2_power, quest3_power, leng2019energy}).

The image sensor's power is dominated by two components: the sensor-host data transfer and the readout circuitry.
Measurements show that transmitting one byte from the image sensor via MIPI CSI-2 interface consumes about 100 \si{\pico\joule} energy~\cite{liu2022augmented} 
which translates to 300 \si{\milli\watt} when transmitting 4K images at 120 FPS.
The readout peripheral circuit that converts the analog pixel value to locally stored digital bits is another dominating component in the image sensor~\cite{choi2015energy}.
While power-efficient ADC design is an active area of research, a survey on the recent image sensors from the past decade~\cite{ma2023leca} shows that the readout circuitry still consumes \hl{66}\% of the sensor's power on average, as shown in \Fig{fig:cis_energy_breakdown}.

\paragraph{Summary.}
Our goal is to significantly reduce the power consumption and latency of eye tracking without hurting the (already sufficient) tracking frequency.
We focus on co-designing the computational image sensor front-end with the eye tracking algorithm, while leaving further optimization of the host SoC hardware to future work.
Optimizing the image sensor not only directly reduces the sensor readout and sensor-host communication power, but also indirectly reduces the amount of work off-sensor algorithms perform.

Hardware acceleration of the off-sensor operations (e.g. segmentation) is orthogonal and complementary to our front-end solution and, thus, out of scope.
In this paper, we assume a standard systolic array architecture to execute any DNNs and claim no novelty for the neural processing unit (NPU) design.

\section{Sparse Sampling-Based Eye Tracking}
\label{sec:algo}

\no{We first describe the sparse sampling algorithm (\Sect{sec:algo:ss}), followed by an eye segmentation algorithm that is robust against sparse inputs (\Sect{sec:algo:dnn}).
We then discuss how two algorithms are jointly trained (\Sect{sec:algo:train}).
This section focuses on the algorithm and leave hardware design to the next section.}

\subsection{Sparse Sampling Algorithm}
\label{sec:algo:ss}

\paragraph{Intuition and Overview.} 
To reduce latency and power consumption, our idea is to perform sparse sampling \textit{inside the sensor}.
In-sensor sampling has two advantages.
First, it reduces the amount of pixels that have to be read out and transferred to the host, two of the main contributors to sensor power consumption.
Second, by reducing the data volume, we also reduce the cost of the downstream eye tracking algorithm.

The overall algorithm pipeline is shown in \Fig{fig:sparse_sample}.
Each frame first gets sparsely sampled by inside the sensor;
the sampled pixels are then transmitted to the host, which executes the eye segmentation and gaze prediction to produce the ultimate gaze information.

Conventional image sampling aims to maximize image reconstruction quality for human vision~\cite{kravets2022progressive, egiazarian2007compressed}.
Instead, our sampling strategy leverages the unique characteristics of, and is thus tailored to, the eye tracking task.
In particular, in eye tracking only the fore-ground eye parts (e.g., pupil, iris, cornea) contribute to the final gaze information.
Naturally, we can approach sampling in two stages: first localizing the fore-ground parts of the eye as the region-of-interest (ROI), followed by sampling within the ROI. 
The ROI prediction DNN is jointly learned with subsequent sampling and downstream eye segmentation to minimize end-to-end loss.

Our two-stage sampling algorithm consists of three serialized stages: eventification, ROI prediction, and sampling.
Let us describe the three stages in detail.



\paragraph{Eventification.}
While one could apply generic, heavy-duty object detection DNNs to detect ROIs (e.g., Mask R-CNN~\cite{he2017mask}), the cost of executing such networks would be prohibitively high for in-sensor computing.

To design a lightweight ROI detection algorithm, the key observation we leverage~\cite{feng2022real} is that in virtually any eye tracking scenario (e.g., AR/VR), the near-eye camera is tightly mounted on the headset, which is in turn tightly mounted on the head. This means the background in eye images is stationary: there is no relative motion between the camera and the eye background. Therefore, any pixel intensity \textit{changes} between consecutive frames inherently indicate the foreground, moving eye parts. The inter-frame pixel differences, thus, provide a natural guide to ROI prediction.

Therefore, the first step in our sampling algorithm is to obtain inter-frame pixel difference, which is expressed as:

\begin{equation}
\label{eqn:eventification}
    E_{t+1} (x, y) = \Phi(|F_{t+1}(x, y)-F_{t}(x, y)|, \sigma)
\end{equation}
\noindent where $F_{t}(x, y)$ and $F_{t+1}(x, y)$ are the pixel values at the $(x, y)$ coordinates at time $T$ and $(T+1)$, respectively; $\Phi$ is an activation function which outputs 1 if the difference is greater than the threshold $\sigma$ (and 0 otherwise). The threshold is a parameter that can be tuned for a specific application or scenario. We empirically find that $\sigma = 15$ yields good results.

The resulting $E$ is essentially a binary event map, where each pixel value indicates whether the corresponding pixel has changed significantly across frames (i.e., an event has occurred) and, thus, belongs to the foreground eye parts.

\paragraph{ROI Prediction.} With the guidance of the event map, we design a lightweight ROI prediction network.
Our ROI prediction network is intentionally small; it contains three convolution (\texttt{Conv}) layers followed by two full-connected (\texttt{FC}) layers, amounting to only $2.1\times 10^7$ MAC operations.
The event map is used as the input to \texttt{Conv} layers.

While event maps are generally effective, there are corner cases where events are not indicative of foreground parts (e.g., blinks, saccades).
To improve the robustness of our ROI prediction, we feed back the segmentation map from the previous frame as a corrective cue similar to prior work~\cite{feng2022real}.

\paragraph{Random Sampling.}
Given an ROI, we randomly sample the pixels inside the ROI.
We find that random sampling is effective even at high sampling rates~(\Sect{sec:eval:sen}).
There are many other sampling alternatives we consider; none works as well.
We will quantitatively compare across different sampling strategies in \Sect{sec:eval:sample}; here we provide an intuitive account.

For instance, one can sample the entire image rather than just the ROI, but the non-ROI regions of an eye image make no contribution to eye tracking result, wasting precious sampling budget.
Alternatively, one can uniformly, rather than randomly, sample the ROI, which would simplify the hardware.
Our results, consistent with prior findings in compressed sensing~\cite{candes2006robust, candes2006stable}, show that uniform sampling significantly reduces the eye segmentation accuracy, suggesting the difficulty of reconstructing the eye parts from uniformly sparse samples.
\no{Finally, one can consider using an additional network to predict which pixels to sample.
The computation cost of doing so is prohibitively high with little accuracy benefit.}

\subsection{Robust Eye Segmentation From Sparse Inputs}
\label{sec:algo:dnn}

Unlike existing eye segmentation algorithms that operate on full eye images~\cite{chaudhary2019ritnet, yiu2019deepvog, kothari2021ellseg, kim2019eye}, our algorithm operates on sparse images (about 5\% of the pixels as \Sect{sec:eval:acc} shows).
Using sparse images means the inputs are susceptible to noise, hurting the accuracy of the algorithm~\cite{dodge2017study}.

Specifically, our experiment (\Sect{sec:eval:acc}) shows that existing Convolutional Neural Network (CNN)-based algorithms often struggle to retain high accuracy at low sampling rates.
\no{CNN-based algorithm's accuracy drops rapidly once the sampling rate (the percentage of pixels that are retained) is below 50\%.}
This is because CNNs inherently rely on local information rather than global information to make predictions.
Consequently, as the sampling rate increases (i.e., fewer pixels), less local information is retained, which leads to accuracy drops.

Instead, we propose a Vision Transformer (ViT)-based segmentation algorithm.
Unlike CNN-based algorithms, ViT leverages an attention-based mechanism that takes into account information from all valid pixels within the input image~\cite{vaswani2017attention}.
Even when the sampling rate is low, ViT still can extract the relationship among pixels that are far away from each other.

\begin{figure}[t]
    \centering
    \includegraphics[width=\columnwidth]{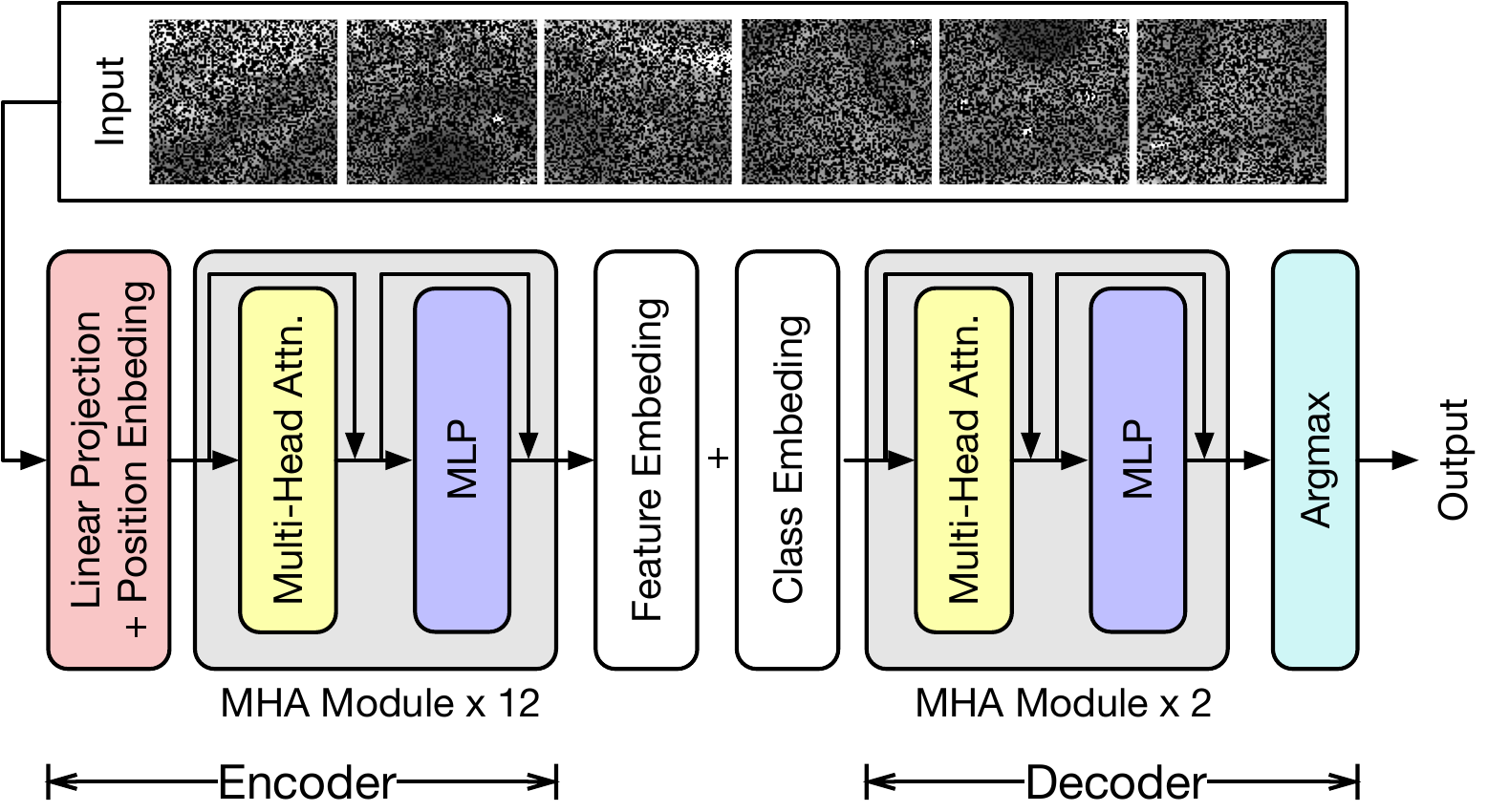}
    \caption{Overall architecture of our ViT segmentation, consisting of an encoder and a decoder. The encoder is composed of a linear projection and 12 MHA modules. The decoder comprises two MHA modules and an argmax layer.}
    \label{fig:vit_arch}
\end{figure}

Our ViT algorithm consists of an encoder and a decoder (\Fig{fig:vit_arch}).
The ViT encoder uses 12 multi-head attention (MHA) modules, similar to the network architecture in Strudel et al.~\cite{strudel2021segmenter}.
Each MHA layer has three heads with a channel size of 192.
To correlate global information, the MLP operations in MHA module compute across all image tokens.
Similarly, ViT decoder uses two MHA layers with the same setting and ends with an argmax layer for the final segmentation prediction.

Note that a ViT network can be readily executed on a typical DNN accelerator (e.g., a systolic array)~\cite{ye2022accelerating, chen2023high}.
Optimizing the accelerator architecture for ViT networks is an active area of reasearch~\cite{wang2022via, lit2022auto};
we leave it to future work to co-design the accelerator with our ViT network.

\subsection{Training Procedure}
\label{sec:algo:train}

The end-to-end tracking algorithm contains two DNNs, one for ROI prediction and the other for ViT segmentation.
We propose a joint training procedure to improve the overall accuracy.
Two loss terms guide our training: a segmentation loss and an ROI loss. The segmentation loss is a cross-entropy loss that governs the accuracy of eye segmentation~\cite{chaudhary2019ritnet}, while the ROI loss uses the conventional mean-square-error loss that governs the ROI prediction accuracy. 
During training, the segmentation loss is back-propagated to both the ROI prediction and the sparse segmentation DNN.
We explicitly mask the gradients belonging to the pixels that are not selected by the random sampling.
That is, only the unmasked gradients are used to update the ROI prediction DNN.

\section{Architectural Support}
\label{sec:sensor}

\no{We first introduce the overall \proj system operation  (\Sect{sec:sensor:pip}) and present an overview of the sensor architecture (\Sect{sec:sensor:overview}), followed by the detailed description of the sensor hardware designs (\Sect{sec:sensor:design}).}


\begin{figure}[t]
    \centering
    \includegraphics[width=\columnwidth]{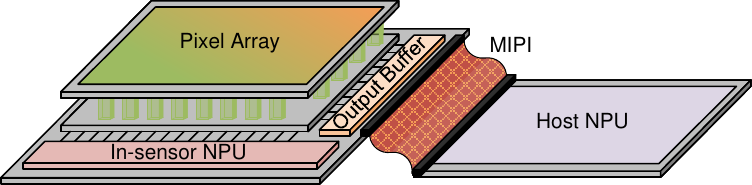}
    \caption{High-level architecture of our eye tracking system.
    The image sensor is connected to the host NPU through the MIPI interface.
    The sensor has a 2-layer DPS pixel array (as with many recent image sensors~\cite{scotttrends}), an in-sensor NPU, and an output buffer.
    Our architectural augmentations lie in augmenting the bottom layer of the pixel array.}
    \label{fig:design}
    \vspace{-5pt}
\end{figure}

\begin{figure*}[t]
\centering
\includegraphics[width=\textwidth]{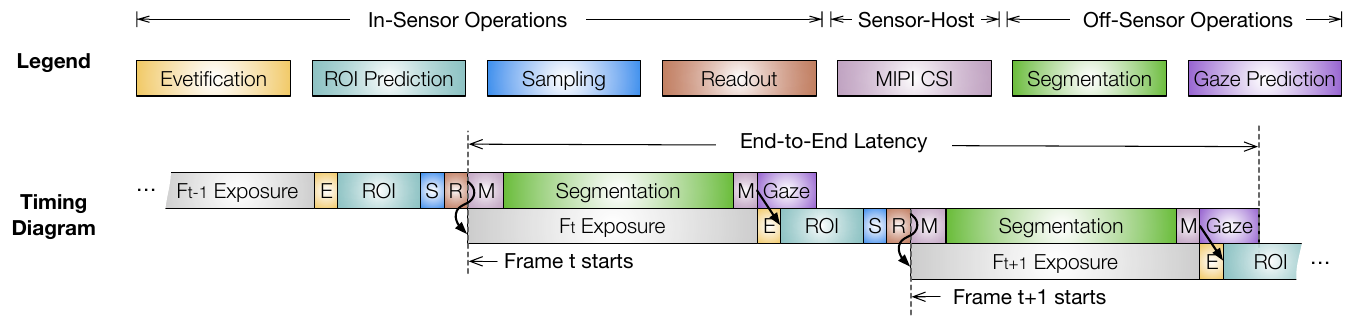}
\caption{Timing diagram of our eye tracking system, which includes operations both in sensor (Exposure, Eventification, ROI Prediction, Sampling, Readout) and off sensor (Eye Segmentation and Gaze Prediction).
We overlap processing of different frames to ensure high frame rate while respecting data dependencies (indicated by filled arrows).
\no{The figure is not drawn to scale; the additional latency introduced by the new in-sensor operations is much smaller compared to the exposure time, so the overall tracking rate is minimally impacted.}
}
\label{fig:pipeline}
\end{figure*}

\subsection{\proj System Overview}
\label{sec:sensor:pip}

We co-design \proj to support the learned sparse sampling algorithm.
The system consists of a computational image sensor and a host NPU connected by the MIPI CSI interface.
The system organization is illustrated in \Fig{fig:design}.
Our main contribution is to architecturally augment the image sensor to support in-sensor sparse sampling (\Sect{sec:algo}) with minimum hardware overhead while leaving the host NPU as is to perform the eye segmentation and gaze estimation tasks.

\Fig{fig:pipeline} depicts the system timing diagram of \proj, which has two main differences when compared with the original eye tracking pipeline in \Fig{fig:overview}.
First, each frame now goes through three additional in-sensor processing stages: eventification, ROI prediction, and in-ROI sampling.
Second, there is a new constraint when pipelining across frames:
$\text{Frame}_t$'s ROI prediction must wait for the segmentation map of $\text{Frame}_{t-1}$ to be sent back from the host via the MIPI CSI interface.
This dependency is purely algorithmic: the previous frame's segmentation map is used as an input to the ROI prediction of the current frame (\Sect{sec:algo:ss}).
The two dependencies are denoted by the arrows in \Fig{fig:pipeline}.

Observing \Fig{fig:pipeline}, it would seem that additional in-sensor operations would increase the eye tracking energy and latency.
As we will describe in the rest of this section, however, the hardware design is such that the additional in-sensor operations introduce negligible latency and energy overhead (2-3 orders of magnitude lower) compared to that of a baseline frame;
meanwhile, in-sensor sampling significantly reduces the data volume involved in readout, MIPI CSI transfer as well as the off-sensor segmentation work which now operates on far fewer pixels. Since both the readout and the MIPI transfer contribute the major sensor energy, and the MIPI transfer contributes the major sensor latency, the sensor's overall energy and latency are notably reduced.

\begin{figure}[t]
    \centering
    \includegraphics[width=\columnwidth]{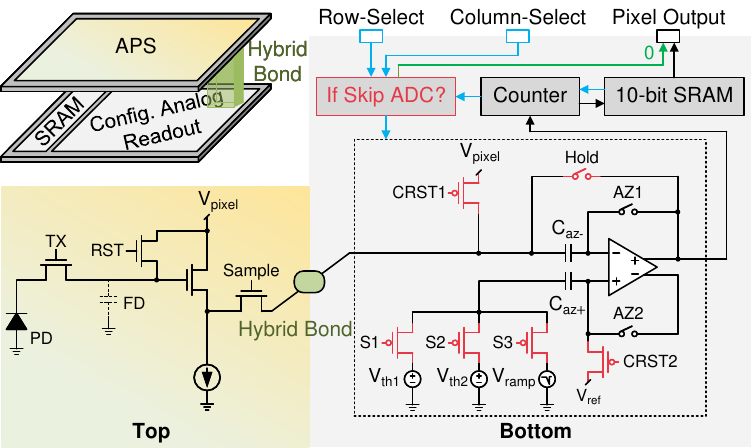}
    \caption{The circuit diagram of the proposed DPS; the red components are new hardware added to a conventional DPS.
    The top layer is a standard pixel design (4T APS) that converts photons to charges, and the bottom layer performs eventification, analog memory, ADC, and sparse readout by reusing the same circuitry.
    Blue arrows: signals used to determine if the pixel performs ADC. Green arrows: signals that output 0 if the pixel skips ADC. Black arrow: output analog pixel value if the pixel performs ADC (sampled).}
    \label{fig:cis_design}
\end{figure}

\begin{figure*}[t]
    \centering
    \includegraphics[width=\textwidth]{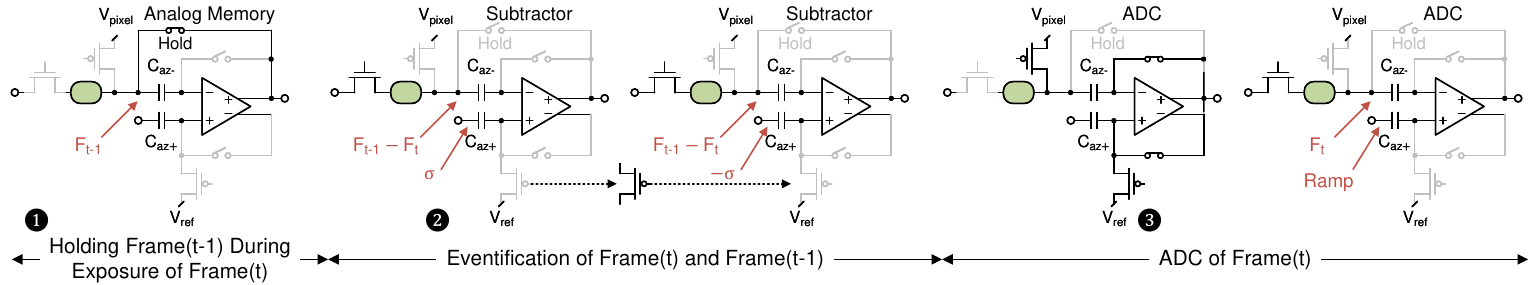}
    \caption{
    Different configurations of the pixel's analog readout circuit. The quantization is only performed for sampled pixels.
    }
    \label{fig:cis_timing}
\end{figure*}

\subsection{Sensor Architecture}
\label{sec:sensor:overview}

The main design consideration for the \proj sensor chip is to support the various in-sensor operations in addition to the normal imaging mode.
Although prior works have introduced designs that could meet one or two of the required operations (see summary in \Sect{sec:related}),
none supports the full gamut of the required in-sensor operations.
Instead of employing dedicated hardware for each function which would incur intolerable area costs at the pixel level, our design principle, is to maximally reuse existing hardware across multiple operations while introducing only minimal augmentation.


As shown in \Fig{fig:design}, the proposed \proj sensor consists of a pixel array, an in-sensor NPU, and an output buffer.
Our hardware augmentation is limited to the pixel array while adopting the standard design strategies for the in-sensor NPU (i.e. systolic array) and the output buffer (i.e. parallel-in-serial-out shift register).
For each frame, the pixel array captures an array of pixels and generates a binary event map (i.e., eventification).
The event map is transferred to the in-sensor NPU, where the ROI prediction DNN resides.
The ROI bounding box is then fedback to the pixel array, which randomly samples pixels in the ROI and reads out only those sampled pixels to the output buffer.
The output buffer connects to the MIPI interface, which transfers the pixels to the host.

The pixel array is implemented following a typical two-layer DPS architecture where the top layer is the array of pixel cells (e.g. a 4-transistor Active Pixel Sensor~\cite{ohta2020smart}), and the bottom layer consists of an array of per-pixel ADC and SRAM. The two layers are connected by per-pixel hybrid bonds. In \Sect{sec:sensor:design}, we will dive into our novel augmentation to each DPS pixel at the bottom logic layer to support additional operations (i.e., eventification and sampling).

\subsection{Design}
\label{sec:sensor:design}

The detailed design of the pixel circuits, which are based on the standard DPS design, is illustrated in \Fig{fig:cis_design}.
For \textit{each} pixel, a standard 4T Active Pixel Sensor (APS) circuit (responsible for converting photons to charges based on the photoelectric effect~\cite{einstein1905molekularkinetischen}) resides in the top layer; the bottom layer contains a 10-bit SRAM and a configurable analog readout circuit.

In a standard DPS, the pixel readout is performed by an Single-Slope ADC (SS ADC) for pixel quantization. The SS ADC operates as follows: a comparator receives the analog pixel value and a monotonically decreasing ramp signal ($V_{ramp}$) at its two input Auto-Zero (AZ) capacitors ($C_{az+}$ and $C_{az-}$), respectively; the comparator's output will not toggle until the ramp signal crosses the analog pixel value; a counter counts the number of cycles it takes for the toggle of the comparator's output, and the counted cycles is the quantized pixel value.

Conventional SS ADC is a fixed-function unit that performs just the quantization.
To perform the in-sensor operations required by our learned sampling algorithm, \proj augments the SS ADC with a few extra switch transistors and a simple logic unit (highlighted in red in \Fig{fig:cis_design}) while reusing many existing ADC components---the AZ capacitors/switches, the comparator, and the counter.
In a sense, we time-multiplex the same analog readout circuit between different operations, e.g., analog buffering, eventification, and normal quantization.
\Fig{fig:cis_timing} shows different configurations of the readout circuit.

We also minimally augment the circuitry to support sparse readout, where only sampled pixels within an ROI go through the ADC and MIPI interface.
This is achieved by the ``If Skip ADC'' logic in \Fig{fig:cis_design}, which is conditioned upon the row-select and column-select signals.
Finally, we reuse the per-pixel SRAM for storing the eventification result and for in-ROI random sampling.
We now discuss the circuit-level behaviors.

\paragraph{Eventification.}
Eventification generates a binary map as the input to the ROI prediction DNN.
According to \Eqn{eqn:eventification}, eventification requires the retention of the previous frame $\text{Frame}_{t-1}$, the subtraction between the current frame $\text{Frame}_{t}$ and the previous frame $\text{Frame}_{t-1}$, and comparing the frame difference with predefined bipolar thresholds $\pm\sigma$.

To hold $\text{Frame}_{t-1}$ during the exposure of $\text{Frame}_{t}$, we configure the comparator as an analog buffer by closing the $Hold$ switch to form a negative feedback loop, as shown in \Fig{fig:cis_timing}~\circled{white}{1}.
$\text{Frame}_{t-1}$ is held on $C_{az-}$, one of the two input AZ capacitors of the analog readout circuit.

The subtraction is done through a switched-capacitor configuration of the comparator circuit.
As shown in \Fig{fig:cis_timing}~\circled{white}{2}, with $Hold$ open to disconnect the negative feedback loop, the charges of the new frame $\text{Frame}_{t}$ are thus transferred onto $C_{az-}$ and the result is naturally the frame difference $(\text{Frame}_{t-1}-\text{Frame}_{t})$.

To compare against a threshold, we utilize the comparison function that an ADC intrinsically performs by simply connecting the other input AZ capacitor, $C_{az+}$, to the thresholding value $\sigma$.
The comparison result is naturally the output of the analog readout circuit, as shown in \Fig{fig:cis_timing}~\circled{white}{2}.
Note, however, that the mathematical formulation (\Eqn{eqn:eventification}) requires comparing against the absolute value of the threshold,
so we apply $\pm\sigma$ sequentially through $V_{th1}$ and $V_{th2}$ (\Fig{fig:cis_design}).

\begin{figure}[t]
    \centering
    \includegraphics[width=\columnwidth]{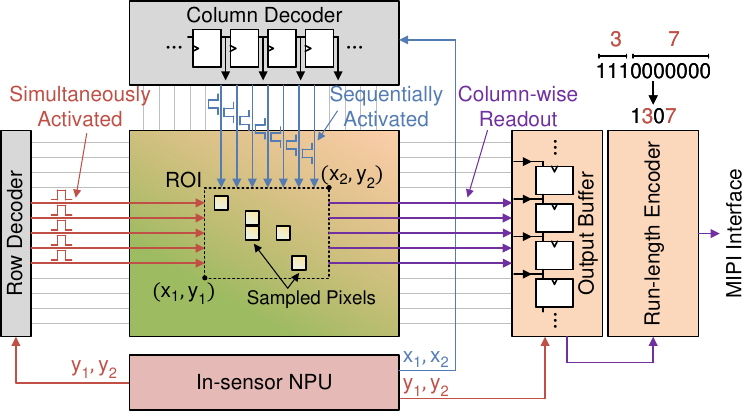}
    \caption{
    The pixel array architecture that allows sparse readout.
    The in-sensor NPU generates the coordinates the two ROI corners ($x_1, y_1$) and ($x_2, y_2$).
    The coordinates are used to driven the row/column decoders to select the ROI.
    All pixels inside the ROI are read to the output buffer (in a column-wise manner),
    but only the sampled pixels are quantized while the unsampled pixels output 0s.
    The output buffer transmits the bits to the MIPI interface through a run-length encoder.}
    \label{fig:roi_select}
\end{figure}

\paragraph{ROI Prediction.}
Following the eventification step, the output of the frame difference compared with the thresholds forms the binary event map and is stored using the per-pixel SRAM.
The SRAM is read by the in-sensor NPU to execute the ROI prediction DNN.
Our design uses a systolic array-like accelerator, and we claim no novelty here (see \Sect{sec:exp}).

The output of the ROI prediction is four numbers ($x_1,x_2,y_1,y_2$), representing the $xy$-coordinates of the two opposing corners of the ROI box.
\Fig{fig:roi_select} shows how these values are used to drive the ROI selection.
The two row coordinates ($y_1,y_2$) are sent to the row decoder and the output buffer, and the two column coordinates ($x_1,x_2$) are sent to the column decoder.
The row decoder activates all the rows between $y_1$ and $y_2$ simultaneously, whereas the column decoder activates all the columns between $x_1$ and $x_2$ \textit{sequentially} because the read-out to the output buffer is necessarily column-by-column.
Note that not all the pixels in the ROI will be read out; only those that are sampled do, the mechanism of which will be discussed next.



\paragraph{Random Sampling.} 
A random bit is generated locally at every pixel to determine whether the pixel will be quantized and read-out.
To avoid additional in-pixel circuitry, we utilize the metastability~\cite{su20071} of the inherent 10-bit SRAM for random bit generation.
The randomness comes from the meta-stability of a typical 6-transistor (6T) SRAM cell when the SRAM is powered-up.
The meta-stable state will randomly latch to a 1 or a 0 due to random noise when the cell is powered-up~\cite{van2012efficient}.
The randomness is not spatially correlated 
due to the differential signaling of the cross-coupled pair.


Although using SRAM for random bit generation requires intermittent SRAM power-up/down, it does not affect the system timing nor the SRAM's memory function (i.e. storing the result of eventification and ADC).
This is because the SRAM is intrinsically power-gated during its inactive periods in the normal functional pipeline: the SRAM is powered-down after the event map is used by the ROI prediction DNN, and will then be powered up to store the quantized pixels.
We leverage the SRAM's intrinsic duty cycle to generate the random bits for every frame.

To control the sample rate,
during a one-time offline calibration all the SRAM cells are powered up and down multiple cycles to profile the distribution of the sum of the 10 power-up bits in each pixel.
From the profiling result, we build a look-up table that translates a sampling rate to a threshold $\theta$.
The $\theta$ is in 4-bit, thus the table has only $2^4$=16 entries to cover the sum which ranges from 0 to 10.
In our simulation, we use the statistics from measurements in prior work ~\cite{holcomb2008power,wieckowski2010black}.

During run time, at each power-up event the counter of each pixel sums the 10 power-up bits by counting the number of 1s in that pixel.
This (4-bit) number is compared with $\theta$ in the ``If Skip ADC'' logic in \Fig{fig:cis_design}.
Only when the sum surpasses $\theta$ will the pixel be actively sampled.
Summing the power-up bits of all the 10 bits in a pixel mitigates the non-uniform randomness across SRAM cells due to the process variation~\cite{holcomb2008power}.


\paragraph{Sparse Readout.}
At this stage, the sampled pixels must be read out.
\Fig{fig:roi_select} shows that all the pixels within the ROI are transferred to the output buffer in a column-wise manner, where the column select signals are sequentially activated (as is in the baseline DPS).
However, only when a pixel is sampled will it be quantized by the ADC, which is controlled by the ``If Skip ADC'' logic in \Fig{fig:cis_design}.
If the pixel is sampled, its comparator is configured to the normal SS ADC (\Fig{fig:cis_timing}~\circled{white}{3}).
If a pixel is not selected, the ``If Skip ADC'' logic connects a constant 0 to the pixel's output port.



The output buffer thus contains both the sampled pixels and the un-selected ones within the ROI.
Since only approximately 20\% of the pixels within the ROI are sampled, we use the run-length encoder~\cite{golomb1966run} to compress the data.
For example, a sequence of 1110000000 is compressed to 1307 where 3 and 7 denote the number of 1s and 0s in the sequence, respectively.
A corresponding run length decoder is implemented in the host NPU to decompress the ROI images before being processed by the eye segmentation algorithm.

\section{Experimental Setup}
\label{sec:exp}

\paragraph{Hardware Configurations.}
The overall hardware system consists of a custom designed image sensor and a conventional DNN accelerator (NPU); we claim no contribution in the latter.
Without losing generality, we assume a systolic array-like NPU, which consist of a $32 \times 32$ MAC array operating at 1 GHz.
This NPU is responsible for computations outside the sensor.
The NPU's global buffer is sized at 2 MB and is banked at a 128 KB granularity.
We also assume a systolic array-style NPU sitting at the bottom layer of the image sensor.
The NPU is consists of an $8 \times 8$ MAC array clocked at 0.5 GHz with a 512 KB SRAM, which is sized to hold the input and intermediate feature maps needed for ROI prediction.

\paragraph{Experimental Methodology.}
All the digital logic is implemented in RTL.
We synthesize, place, and route the design using an EDA flow consisting of Synopsys and Cadence tools.
The SRAMs are compiled by an ARM memory compiler.
Power is simulated using Synopsys PrimeTimePX, with fully annotated switching activity.
The pixel design on the sensor top layer follows that in Seo et al.~\cite{seo20222}.
The analog circuit on the bottom layer is implemented in standard CMOS 65 \si{\nano\meter} technology and simulated using Cadence Virtuoso.

\begin{figure*}[t]
  \centering
  \captionsetup[subfigure]{width=0.98\columnwidth}
  \subfloat[\small{Vertical angular error.}]
  {
  \includegraphics[width=.98\columnwidth]{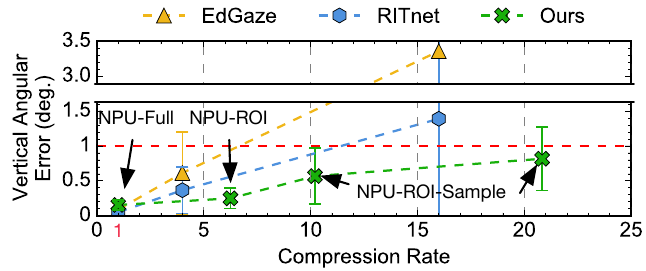}
  \label{fig:vertical_error}
  }
  \subfloat[\small{Horizontal angular error.}]
  {
  \includegraphics[width=.98\columnwidth]{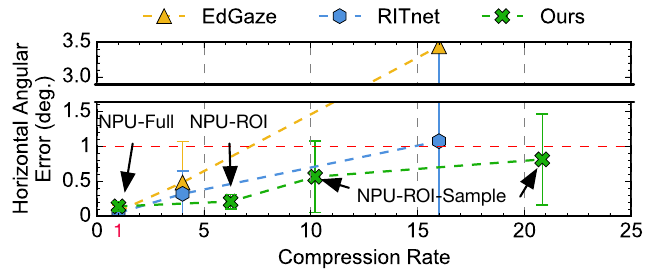}
  \label{fig:horizontal_error}
  }
  \caption{End-to-end gaze prediction vs. compression rate (uncompressed size over compressed size; 1 for full frame). \hl{The error bars denote one standard deviation. We annotate different variants of our method for ablation studies. \textbf{NPU-Full} opeates on full eye images. \textbf{NPU-ROI} applies operates on ROI images. \textbf{NPU-ROI-Sample} is our full-fledged pipeline.}}
  \label{fig:acc}
\end{figure*}

Following the typical technology nodes in today's image sensors, we assume that the top layer of the image sensor uses a standard CMOS 65 \si{\nano\meter} process node, the bottom analog and logic layer uses a 22 \si{\nano\meter} process node, and the off-sensor NPU uses a 7 \si{\nano\meter} process node.
\hl{We use the synthesis results from a TSMC 16 \mbox{\si{\nano\meter}} FinFET library and scale the results to other process nodes using the DeepScaleTool~\mbox{\cite{stillmaker2017scaling, sarangi2021deepscaletool}}, which models the classic CMOS scaling by ``\textit{fitting published data by a leading commercial fabrication company for silicon fabrication technology generations from 130 \mbox{\si{\nano\meter}} to 7 \mbox{\si{\nano\meter}}.}'' }

We model noises in the image sensor, following classic analytical models of various noise sources~\cite{rowlands2017physics, farrell2012digital, brooks2019unprocessing, ohta2020smart}.
Specifically, the analog readout circuits (on the bottom layer of the sensor) are carefully designed such that its read noise does not introduce functional errors to the binary eventification and ADC quantization.
We model the photon shot noise using the classic method (drawing from a Poisson distribution)~\cite{ohta2020smart} and considered it during training and evaluation.

The DRAM parameters are modeled after Micron 16 Gb LPDDR3-1600 (4 channels) as detailed in its datasheet~\cite{micronlpddr3}.
The calculation of DRAM energy is based on Micron's System Power Calculators~\cite{microdrampower} using the memory traffic, including kernels and activations of segmentation ViT.
We use the energy per byte over the MIPI CSI interface from Liu et al.~\cite{liu2022augmented}.



\paragraph{Algorithm Baselines.}
To evaluate the accuracy of our ViT-based eye segmentation algorithm (specifically designed to leverage the sparse eye image input; \Sect{sec:algo:dnn}), we compare against two state-of-the-art eye segmentation algorithms, both operate on dense eye images: \mode{RITnet}~\cite{chaudhary2019ritnet}, which uses an encoder-decoder architecture, and \mode{EdGaze}~\cite{feng2022real}, which uses depthwise separable convolution network.  


We follow the same training procedure in prior work~\cite{feng2022real, kothari2021ellseg} and use OpenEDS~\cite{garbin2019openeds}, a widely-used eye tracking dataset.
We train the eye segmentation algorithm using a batch size of 4 with 250 epochs.
We train the ROI prediction network for 100 epochs with a batch size of 8.
 
\paragraph{System Variants.}
\hl{To tease apart the contribution of different components in our system and to support ablation studies}, we compare against the following variants:
\begin{itemize}
    \item \mode{NPU-Full}: represents a conventional eye tracking system: a non-computational image sensor with a host NPU. The sensor transmits the full-size eye images to the host NPU, which executes the eye segmentation algorithm.
    \item \mode{NPU-ROI}: this variant has the same hardware configuration as \mode{NPU-Full}, except the host NPU executes the ROI prediction DNN to extract the ROI, on which the subsequent eye segmentation algorithm operates.
    \item \mode{S+NPU}: same as our proposed design except it executes sparse sampling in the digital domain inside the sensor.
\end{itemize}

\section{Evaluation}
\label{sec:eval}

This section starts by demonstrating that \proj achieves adequate accuracy against baselines even when significantly reducing the pixels (\Sect{sec:eval:acc}). Following this, we demonstrate that our sensor design reduces the overall energy consumption (\Sect{sec:eval:energy}) and tracking latency (\Sect{sec:eval:perf}).
We show that our hardware augmentation introduces little area overhead (\Sect{sec:eval:area}) and that our sampling strategy out-performs alternatives (\Sect{sec:eval:sample}).
Finally, we conduct a sensitivity study to understand how \proj's performance and energy savings vary under diverse settings (\Sect{sec:eval:sen}).

\subsection{Accuracy vs. Compression Rate}
\label{sec:eval:acc}


Our eye tracking algorithm achieves higher accuracy compared to existing eye tracking algorithms across a range of compression rates (uncompressed size over compressed size).
\Fig{fig:acc} presents the accuracy-vs-compression-rate comparisons on both vertical angular error (\Fig{fig:vertical_error}) and horizontal angular error (\Fig{fig:horizontal_error}).
The input images to the two baseline algorithms are downsampled by different amounts to achieve different compression rates.

Across all compression ratios, our algorithm consistently maintains the gaze estimation accuracy within the acceptable error threshold ($1^{\circ}$) in both directions\hl{\mbox{\cite{smarteyetracker, tobiieyetracker, eyelink, htc_eye_tracker}}}.
Specifically, we achieve a $20.6\times$ data reduction with $0.8^{\circ}$ vertical angular error and $0.7^{\circ}$ horizontal angular error.
Unless otherwise noted, this is the compress rate we will use in the rest of the evaluation.
Our algorithm also consistently outperforms existing algorithms across all compression rates \hl{with much higher robustness.
The robustness can be seen by comparing the standard deviation of our method with that of the two baselines: our method has a much smaller accuracy variation, showing a stronger ability to tolerate temporal drifts.} 
While not shown in the figure, our algorithm is also more computationally efficient compared to \mode{RITnet} and \mode{EdGaze}.
For instance, compared to \mode{RITnet}, we reduce the MAC operation counts by a factor of 4.



\subsection{Energy Reduction}
\label{sec:eval:energy}

\begin{figure}[t]
    \centering
    \includegraphics[width=\columnwidth]{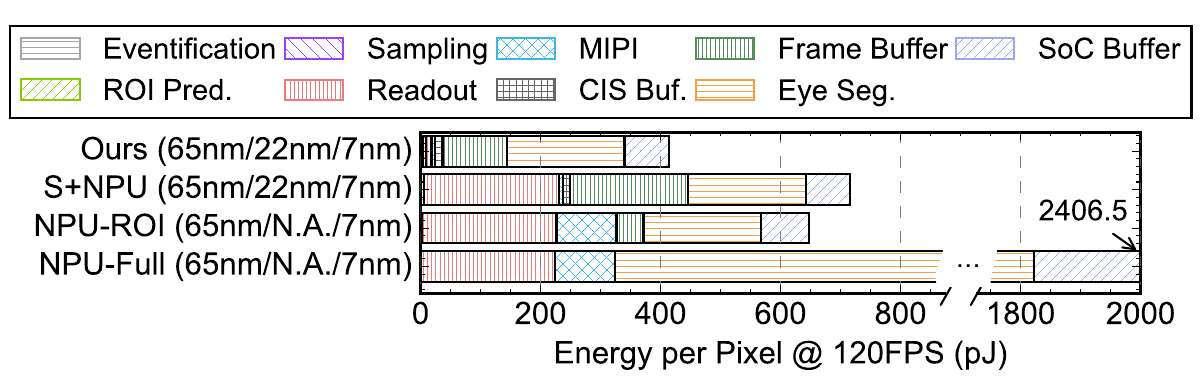}
    \caption{Comparison of energy savings across different sensor-SoC designs at 120 FPS. Numbers inside each parenthesis represent the process node of sensor analog layer, sensor digital logic layer and host SoC, respectively.}
    \label{fig:energy}
\end{figure}

\proj also significantly reduces the eye tracking energy consumption.
\Fig{fig:energy} compares \proj with three variants disucssed as the end of \Sect{sec:exp} using a stacked bar plot, dissecting the contributions of different hardware components.


Compared to \mode{NPU-Full}, \proj achieves a 4.0$\times$ energy reduction.
The reduction comes from three sources: the analog readout energy, MIPI data transfer energy, and the off-sensor work (e.g., eye segmentation NPU and accessing the on-chip buffer).
The latter is especially significantly, contributing to 60.1\% energy of \mode{NPU-Full}.
By predicting ROIs and operating only on sampled pixels, \proj reduces the off-sensor work significantly.


While performing ROI prediction reduces overall energy, where ROI prediction is executed also matters.
\mode{NPU-ROI} executes ROI prediction on the host SoC, taking advantage of the more advanced process nodes, thus reducing the energy spent on executing ROI prediction.
In contrast, \mode{S+NPU} executes ROI prediction inside the sensor, which has the advantage of reducing readout and MIPI energy but increases the ROI prediction and buffer energy, because the process node of the sensor uses an older process node.
As a result, \mode{S+NPU} actually increases the energy by 1.1$\times$ over \mode{NPU-ROI}, mainly due to the high leakage power of the in-sensor frame buffer.
The leakage power of the frame buffer can not be eliminated by power gating because the frame buffer must continuously retain the previous frames for eventification.

\proj combines the benefits of both \mode{S+NPU} and \mode{NPU-ROI} by storing the previous frames in analog memory and executes eventification in the analog domain.
That way, \proj reduces both the in-sensor frame buffer energy and the MIPI data transfer.
As a result, \proj achieves 1.7$\times$ and 1.6$\times$ energy reduction compared to \mode{S+NPU} and \mode{NPU-ROI}, respectively.

\paragraph{Overhead.}
The results above show that the overhead introduced by \proj is clearly out-weighted by its benefits.
In particular, there are two main energy overhead: the additional transfer of the previous frame's segmentation map from the host SoC (in assisting ROI prediction; see \Fig{fig:pipeline}) and the RLE (see \Fig{fig:roi_select}).
The two sources account for only 0.6\% and 0.04\% of the overall energy, respectively.

\subsection{Tracking Latency and Frequency}
\label{sec:eval:perf}

The energy saving of \proj comes with little impact on the overall tracking frequency while significantly reducing the tracking latency.
To ensure a fair comparison, we choose the same process node combination across all variants,
and the clock rates of sensors and host SoCs are set to 0.5~\si{\giga\hertz} and 1~\si{\giga\hertz}, respectively, across all variants.

\begin{figure}[t]
    \centering
    \includegraphics[width=\columnwidth]{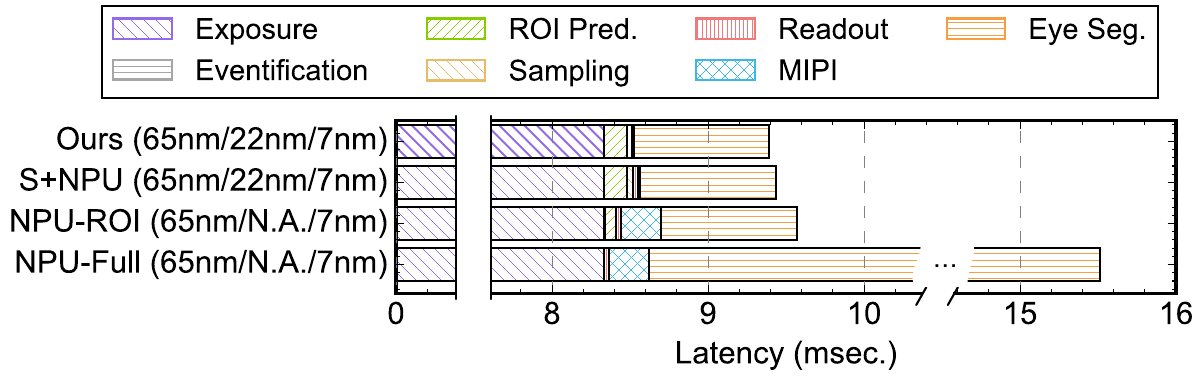}
    \caption{End-to-end latency comparison across different sensor-SoC designs at 120 FPS. Our sensor design does not affect the sensor exposure and achieves similar latency as \mode{S+NPU} and \mode{NPU-ROI}. Numbers inside each parenthesis represent the process node of sensor top layer, sensor bottom logic layer and the host SoC, respectively.}
    \label{fig:latency}
\end{figure}


\Fig{fig:latency} compares the end-to-end tracking latency under the 120 FPS requirement.
\proj reduces the tracking latency by 1.4 $\times$ over \mode{NPU-Full}, primarily because the segmentation DNN latency is accelerated by 7.7$\times$, since it operates only on a small amount of pixels \hl{(10.8\%)}. \hl{The average execution time of eye segmentation is 0.87 \mbox{\si{\milli\second}} with a standard deviation of 0.48 \mbox{\si{\milli\second}}. 
The latency varies across frames, because different frames have different ROI sizes: the average ROI size is 34257.8 pixels with a standard deviation of 18803.6.}

Even with additional work in the analog domain, our latency is similar to that of \mode{S+NPU} and \mode{NPU-ROI}.
This is because the latency in all three schemes is by far dominated by the exposure time, which is held constant.
The additional computations introduce a latency overhead that is orders of magnitude shorter than the exposure time.
For instance, compared to a 8.3 \si{\milli\second} exposure time, eventification and ROI prediction introduce an overhead of 5 \si{\micro\second} and 150 \si{\micro\second}, respectively.


Because the in-sensor analog operations are much faster than the exposure time, \proj also has little effect on the exposure time (see \Fig{fig:pipeline}).
Overall, \proj reduces the overall exposure time by only 1.8\%, which has a minimal impact on the overall eye tracking accuracy (as results in \Fig{fig:acc} factor in this exposure time change).


\subsection{Area Estimation}
\label{sec:eval:area}

While the power and the timing of the proposed DPS are directly obtained from circuit simulations, the area has to be estimated: DPS consists mostly of analog circuitry whose area is sensitive to manual layout and, thus, is not directly available from synthesis.
Our pixel area estimation is based on previous DPS designs that have similar complexity.

Specifically, compared to a typical DPS with the ADC function only, our hardware augmentation to support functional multiplexing takes up only 7 extra switching transistors and simple digital logic (red components in \Fig{fig:cis_design}), whose area is estimated to be comparable to 12 single-bit SRAM cells.
The bottom layer of our design has 2 capacitors (233 fF each), one comparator, 13 switching transistors, 10 6T SRAM cells, and trivial digital logic (a 4-bit digital comparator, 21 gates) in 22~\mbox{\si{\nano\meter}} technology.

Comparably, a stacked DPS by Meta~\cite{ikeno20224} has 2 capacitors, one comparator, 28 switching/logic transistors, and 10 6T SRAM cells on its bottom layer, achieving 4.6$~\si{\micro\meter}$ pixel size in 65~\si{\nano\meter} node.
Another stacked DPS by Samsung~\cite{seo20222} has one comparator, one positive-feedback amplifier, and 22 6T SRAM cells on its bottom layer, achieving 4.95$~\si{\micro\meter}$ pixel size in 28~\si{\nano\meter} node.
\hl{Thus, we choose a pixel pitch of $5~\mbox{\si{\micro\meter}} \times 5~\mbox{\si{\micro\meter}}$.}

\hl{With this pixel size, the pixel array (\mbox{$640\times 400$}), the in-sensor NPU, and the output buffer (including run-length encoder) attribute to 6.4~\mbox{\si{\milli\meter^2}}, 0.4~\mbox{\si{\milli\meter^2}}, and 0.1~\mbox{\si{\milli\meter^2}}, respectively.}
The run-length decoder area on the host is also negligible; it is estimated to be less than 0.1\% of the host area.

\subsection{Comparison with Sampling Alternatives}
\label{sec:eval:sample}


We show that the in-ROI pseudo-random sampling strategy outperforms other sampling alternatives:
\begin{itemize}
    \item \mode{Full+Random}: a method that uniformly at random samples the full-size frame without ROI prediction
    \item \mode{Full+DS}: a method that uniformly downsamples the full-size frame without ROI prediction
    \item \mode{Skip}: a method that detects the event density within each frame to determine whether to skip subsequent eye segmentations and reuse previous segmentation results~\cite{feng2022real}
    \item \mode{ROI+DS}: a method that uniformly downsamples within the predicted ROI
    \item \mode{ROI+Fixed}: a method that uses dataset statistics to overfit a fixed ROI sampling mask offline
    \item \mode{ROI+Learned}: a method that uses an additional ViT network to learn the pixel sampling within the ROI
\end{itemize}

\begin{figure}[t]
  \centering
  \includegraphics[width=\columnwidth]{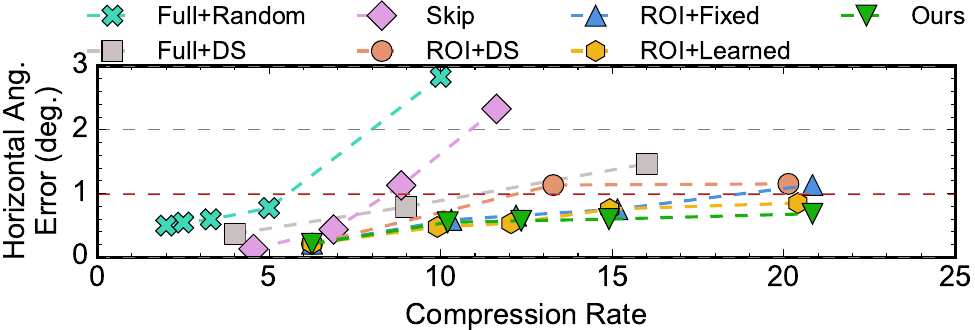}
  \caption{Comparison between our sparse sampling and other sampling alternatives. Our method can retain acceptable accuracy even at high compression rates.
  }
  \label{fig:sampling_alternative}
\end{figure}

\Fig{fig:sampling_alternative} compares the horizontal angular error under different compression rates.
\proj consistently outperforms all other methods across all compression rates.
The highest accuracy gains are achieved against the full-frame methods, showing the benefits of ROI prediction.
At a 21$\times$ compress rate, only ours and \mode{ROI+Learned} can achieve an accuracy below the tolerable threshold of $1^{\circ}$.
\mode{ROI+Learned}, however, requires an additional in-sensor DNN to predict the sampling pattern, introducing intolerable pixel-wise overhead.


\subsection{Sensitivity Study}
\label{sec:eval:sen}

\begin{figure}[t]
\centering
\begin{minipage}[t]{0.48\columnwidth}
  \centering
  \includegraphics[width=\columnwidth]{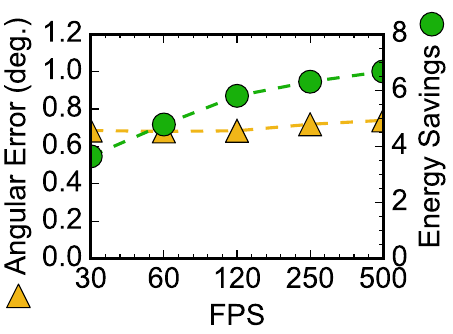}
  \caption{Sensitivity of end-to-end gaze error and energy saving over \mode{NPU-Full} with respect to frame rate.}
  \label{fig:fps_sen}
\end{minipage}
\hspace{2pt}
\begin{minipage}[t]{0.48\columnwidth}
  \centering
  \includegraphics[width=\columnwidth]{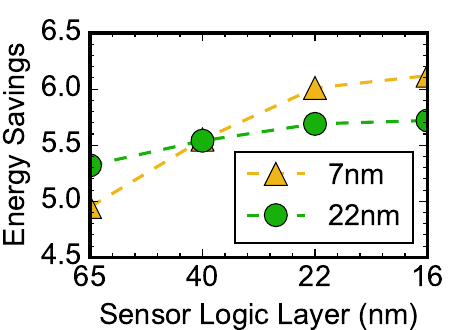}
  \caption{Energy saving over \mode{NPU-Full} with respect to logic layer's process node under two SoC process nodes.}
  \label{fig:energy_profile}
\end{minipage}
\end{figure}


\paragraph{Frame Rate.}
\Fig{fig:fps_sen} shows how the end-to-end horizontal gaze accuracy (left $y$-axis) and energy saving over \mode{NPU-Full} (right $y$-axis) change with the prescribed frame rate.
The overall gaze error slightly increases by 0.03$^{\circ}$ as the frame rate increases from 30 FPS to 500 FPS.
The primary contributor to the accuracy drop is that a higher frame rate reduces exposure time, which leads to a lower SNR (primarily driven by the photon shot noise~\cite{rowlands2017physics}). 
Nevertheless, \proj retains tolerable gaze accuracy (1$^{\circ}$) even at 500 FPS.
As the frame rate increases from 30 FPS to 500 FPS, the energy saving over \mode{NPU-Full} increases from 3.6$\times$ to 6.7$\times$.
The energy reduction is because a higher frame rate means shorter exposure time, which reduces the retention time of the analog frame buffer and reduces the leakage energy.

\paragraph{Process Node.} 
\hl{Throughout the sensitivity study we keep the process nodes for both the top layer and the bottom analog circuits fixed; and we synthesize the logic layer and off-sensor NPU with a TSMC 16 \mbox{\si{\nano\meter}} FinFET library, and use DeepScaleTool\mbox{\cite{stillmaker2017scaling, sarangi2021deepscaletool}} to scale them to different nodes.
We note that mixing technology nodes \textit{is} the norm in image sensors, as discussed in \mbox{\Sect{sec:motiv:cis}}.
}

\Fig{fig:energy_profile} shows the sensitivity of the energy saving (over \mode{NPU-Full}) as the sensor and SoC process nodes change.
We study two SoC process nodes (the two curves), 7 \si{\nano\meter} and 22 \si{\nano\meter}, and sweep (on the $x$-axis) the process node of the sensor's logic layer from 16 \si{\nano\meter} to 65 \si{\nano\meter}.

The overall energy saving is more sensitive to the processor node of the logic layer when the SoC uses a 7 \si{\nano\meter} node over that of a 22 \si{\nano\meter} node.
This reason is simple: when the SoC uses a 22 \si{\nano\meter} node, the off-sensor work tends to dominates the total energy, leaving less room for optimization.

\begin{table} 
\centering
\caption{\hl{Sensitivity of gaze error, standard deviation, and energy savings (over \mbox{\mode{NPU-ROI}}) to ROI reuse window.}}
\resizebox{.95\columnwidth}{!}{
\renewcommand*{\arraystretch}{1}
\renewcommand*{\tabcolsep}{5pt}
\begin{tabular}{ ccccc } 
\toprule[0.15em]
\textbf{\specialcell{Reuse Window}} & 1 & 4 & 16 \\ 
\midrule[0.05em]
\textbf{\specialcell{Vertical Angular Error (std.)}} &  0.25 (0.15) & 0.49 (0.30) & 0.75 (0.69) \\
\midrule[0.05em]
\textbf{\specialcell{Energy Savings}} & 0\%
 & 0.023\% & 0.029\% \\
\bottomrule[0.15em]
\end{tabular}
}
\label{tab:roi_reuse}
\end{table}

\hl{\mbox{\paragraph{ROI Reuse.}} 
Instead of predicting an ROI for each frame, one can also reuse a previously ROI.
We implement a ROI-reuse version of \mbox{\mode{NPU-ROI}}, where a previously predicted ROI is reused over a number of subsequent frames quantified by the \textit{reuse window}.
\mbox{\Tbl{tab:roi_reuse}} shows how the gaze error and energy savings of the ROI-reused version of \mbox{\mode{NPU-ROI}} over itself without ROI reuse change with the reuse window.

Overall, reusing previous ROIs leads to a significant accuracy drop with negligible energy savings.
For instance, when we reuse an ROI for the next consecutive 16 frames, there is only a 0.03\% energy saving but an $0.75^{\circ}$ error increase.
Worse, the standard deviation of the angular error increases with the reuse window, showing a decrease in robustness.
The reason the energy gain is small from ROI reuse is because the prediction network’s energy consumption is insignificant (1.04\% of the total in-sensor energy).
}



\section{Related Work}
\label{sec:related}


\paragraph{Computational Image Sensors.}
Image sensors are increasingly integrating computation capabilities.
The computation is conventionally done in the digital domain, such as Sony IMX 500 CIS~\cite{eki20219}, which integrates DNN accelerator inside the sensor.
Recent proposals move computation into the analog domain to, e.g., computes logarithmic gradients~\cite{young2019data}, extract HOG features~\cite{ma2022hogeye}, eventification~\cite{lichtsteiner2008128, feng2023learned}, random sampling~\cite{oike2012cmos}, ROI-based readout~\cite{choi2007spatial}, and even analog DNNs~\cite{choi2020design, xu2021senputing}.
\proj performs eventification, random sampling, and ROI-based readout inside the sensor with little area overhead by reusing existing in-sensor hardware.

\hl{RedEye\mbox{\cite{likamwa2016redeye}} is proposed as an analog ConvNet image sensor architecture.
\mbox{\proj} exhibits three major differences. 
First, most importantly, the two works differ in high-level system design strategies: RedEye splits pre-trained DNNs to execute early DNN layers inside the sensor.
It does not consider co-design or joint training of the algorithm with the hardware and is highly constrained in the type and size of the DNN layers it can accommodate. In contrast, \mbox{\proj} co-designs the in-sensor operations (sparse sampling) with the off-sensor downstream DNN, and therefore works more flexibly with diverse downstream vision algorithms and network architectures.
Second, they have different hardware implementations: RedEye assumes a conventional analog image sensor and implements the DNN layers in the analog domain whereas \mbox{\proj} adopts a stacked DPS, where the DNN executes in the digital layer, naturally mitigating the noise issue in analog processing.
Finally, \mbox{\proj} also proposes in-sensor sparse sampling and readout, which are unconcerned with in RedEye.
}

\proj not only augments the sensor hardware but jointly designs the in-sensor work (sampling) with off-sensor algorithm (segmentation) to ensure high task accuracy.
Prior work explored such co-designs. LeCA~\cite{ma2023leca} jointly trains an in-sensor encoder with downstream tasks; Bong et al. construct image sensor-processor systems for eye tracking~\cite{bong20160} and face recognition~\cite{bong201714}, respectively; their eye tracking system only achieves 30~FPS and consumes 4.3~\si{\nano\joule} per pixel, which is more than 10$~\times$ higher than that of \proj.

\hl{
\mbox{\paragraph{Random Sampling Image Sensors.}}
The concept of random sampling in image sensors has been explored in compressive sensing and HDR imaging applications. In compressive sensing, random numbers are spatially assigned to the pixel array. However, the random number generator is either off-chip in the optical domain\mbox{\cite{duarte2008single}} or on-chip but resides beside the pixel array for row-wise processing\mbox{\cite{oike2012cmos}} or requires complicated in-pixel logic\mbox{\cite{majidzadeh2010256}}. Thus, none meets the requirement for compact DPS in \mbox{\proj}. In HDR imaging, random numbers are temporally assigned to each pixel’s sub-exposure slots. However, the random numbers are generated with coarse granularity (in pixel tiles)\mbox{\cite{yoshida2019high}} or complicated in-pixel logic\mbox{\cite{martel2020neural}}. More crucially, such coded exposure scheme destroys the original pixel value, making it unsuitable for our design.

In contrast, \mbox{\proj} exploits existent in-pixel SRAMs with simple logic to realize fine-granular random sampling with a compact design, and buffer the necessary pixel values for eventification.
The sampling method implemented in our sensor is customized to eye tracking, but prior work\mbox{\cite{liu2023patchdropout}} has shown that other computer vision tasks such as classification can also benefit from sparse sampling.  While our sampling network will have to adapt to different tasks, the actual hardware support for random sampling (\mbox{\Fig{fig:roi_select}}) readily applies.
}



\paragraph{Eye Tracking Acceleration.} 
Researchers explored dedicated accelerators to accelerate eye tracking.
Bong et al.~\cite{bong20160} and Hong et al.~\cite{hong20152} design accelerators for in-sensor gaze estimation. They target non-DNN algorithms with much inferior accuracy as compared to the state of art DNN-based algorithms.
I-flatcam~\cite{zhao2022flatcam} and EyeCoD~\cite{you2022eyecod} design accelerators for an eye tracking algorithm targeting lensless cameras while leaving the sensor front-end un-optimized.
\proj shows that reducing sensor readout and sensor-host communication leads to significant overall energy reduction.

Prior studies have explored lightweight eye segmentation algorithms~\cite{feng2022real, you2022eyecod, zhao2022flatcam, ma2023camj}.
For instance, EdGaze~\cite{feng2022real} predicts the ROI of an image before segmentation.
Previous work has also explored ROI-based machine vision systems~\cite{ma2023camj, feng2023learned, kodukula2021rhythmic}.
Built on top of the ROI prediction idea, we show that pseudo-random sampling within the ROI can further reduce energy consumption with minimal hardware support.
We also co-design a ViT-based segmentation DNN to be robust against sparse inputs, as the accuracy of previous eye segmentation algorithms tends to drop under sparse inputs.

\hl{\mbox{\paragraph{Event Cameras.}}
Readers familiar with event cameras~\mbox{\cite{gallego2020event}} might recognize that our eventification algorithm (\mbox{\Eqn{eqn:eventification}}) is an emulation of an event camera.
Indeed, our idea is inspired by event cameras
--- with one crucial difference: classic event cameras normalize pixel difference with respect to the previous value.
We simplify the design to remove the normalization operation, which only complicates the sensor hardware design without providing noticeable accuracy benefits for eye tracking as we empirically find.

While there are generic object/ROI detection algorithms in event cameras\mbox{\cite{mitrokhin2018event,perot2020learning}}, our lightweight ROI detection can be seen as a specialized algorithm tailored for eye tracking, based on the observation that background pixel values in eye tracking do not change much over frames, so a very lightweight frame differencing would reveal ROI.}

\section{Conclusion}
\label{sec:conc}

Sparse in-sensor sampling is critical to reducing the energy consumption and end-to-end latency of eye tracking, a crucial component in emerging domains such as AR/VR.
Sampling within an image sensor dramatically reduces the amount of data that has to go through the energy-intensive image readout chain and sensor-host communication interface, two major power contributors to image sensors.
The host, as a result, also operates on far fewer pixels, further reducing the computation cost.
To support in-sensor sparse sampling with little hardware augmentation, \proj reuses the existing pixel-level analog readout circuitry to support eventification, random sampling, and sparse readout; and \proj uses a small in-sensor NPU to support ROI prediction. 
\proj reduces pixel volume by about 95\% and thus achieves an $8.2~\times$ energy reduction and a $1.4~\times$ tracking latency reduction with little tracking accuracy degradation, compared to existing eye tracking systems.

\section{Acknowledgement}

We thank anonymous reviewers from ISCA 2024 for their valuable comments.
The work is partially supported by NSF Awards \#2328856, \#2416375, \#1942900.


\bibliographystyle{IEEEtranS}
\interlinepenalty=10000
\bibliography{refs}

\end{document}